\documentclass{nature}

\usepackage{xurl}
\usepackage{times}

\usepackage[english]{babel}
\usepackage[utf8]{inputenc}
\usepackage[T1]{fontenc}
\usepackage{lineno}
\usepackage{pdfpages}
\usepackage[version=4]{mhchem}
\makeatletter
\usepackage[margin=2.54cm]{geometry}

\usepackage{lipsum}
\usepackage{mhchem}
\usepackage{amsmath}
\usepackage{graphicx}
\usepackage[colorlinks=true, allcolors=blue]{hyperref}
\usepackage{caption}
\title{Optically induced thermal demagnetization and switching of antiferromagnetic domains in NiO and CoO thin films}

\author{Maciej D\k{a}browski$^1\dagger$, Tong Wu$^2$, Connor R. J. Sait$^1$, Jia Xu$^3$, Paul S. Keatley$^1$, Yizheng Wu$^{2,4}$ Robert J. Hicken$^1$, Olena Gomonay$^{5\dagger}$, }

\makeatletter
\let\saved@includegraphics\includegraphics
\AtBeginDocument{\let\includegraphics\saved@includegraphics}
\renewenvironment*{figure}{\@float{figure}}{\end@float}
\makeatother

\begin{document}

\maketitle

\begin{affiliations}
\item Department of Physics and Astronomy, University of Exeter, EX4 4QL, Exeter, UK
\item Department of Physics and State Key Laboratory of Surface Physics, Fudan University, Shanghai 200433, China
\item Department of Physics, School of Physics and Telecommunication Engineering, Shaanxi University of Technology, Hanzhong 723001, China
\item Shanghai Key Laboratory of Metasurfaces for Light Manipulation, Fudan University, Shanghai 200433, China
\item Institute of Physics, Johannes Gutenberg-University Mainz, D-55099, Mainz, Germany

$^{\dagger}$Corresponding authors: m.k.dabrowski@exeter.ac.uk and ogomonay@uni-mainz.de

\end{affiliations} 
\clearpage
\begin{abstract}

\section*{Abstract}

We demonstrate all-optical manipulation of magnetic domains in NiO/Pt and CoO/Pt thin films with insulating antiferromagnetic layers. Using magneto-optical birefringence imaging, we show that even a single laser pulse can thermally demagnetize the antiferromagnet, leading to a random redistribution of domains. By sweeping the laser beam, controlled domain wall motion is induced, enabling partial switching of the antiferromagnetic order. The behavior is captured by an analytical model in which temperature gradients generated by the moving beam exert a thermal pressure on domain walls in the form of a ponderomotive force. Importantly, the 90$^{\circ}$ domains can be reversibly toggled solely by reversing the direction of the thermal gradient, demonstrating all-optical switching without the need for electric currents. These findings establish a route toward ultrafast optical manipulation of fully compensated antiferromagnets, with potential impact on non-volatile memory technologies and antiferromagnetic spintronics.

\end{abstract}
\setlength{\footskip}{15pt}

\section*{Introduction}
Over the past decade, lasers have been successfully used to control the spin of an electron via optically driven thermal and electronic excitations. Most notably, all-optical switching (AOS) of the magnetization was demonstrated for various ferromagnetic and ferrimagnetic
materials \cite{Stanciu2007,Lambert2014,Stupakiewicz2017,Banerjee2020,Dabrowski2021,Igarashi2023,Davies2024}, which is of great interest for magnetic recording \cite{Pezeshki2024}. In general, AOS can be either helicity-dependent or helicity-independent, and can be achieved with single or multiple laser pulses. Optical pumping of magnetic layers is highly sensitive to the laser parameters, such as fluence and the number of pulses, and often results in so-called thermal demagnetization. Unlike AOS, thermal demagnetization leads to a random redistribution of magnetic domains. Regardless of the outcome, magnetic domain imaging remains the most effective experimental method for demonstrating and optimizing AOS and related phenomena \cite{Hadri2017}.

In contrast to ferromagnets and ferrimagnets, optical control of antiferromagnets (AFMs) remains largely unexplored \cite{Manz2016,Grigorev2022,Meer2023}, mainly due to the difficulty of detecting AFM domains arising from the absence of net magnetization and stray fields. Nevertheless, AFMs are increasingly recognized as promising candidates for future energy-efficient, high-speed, high-density spintronic and data storage applications \cite{Wadley2016,Hou2019,Jungfleisch2018,Kampfrath2010,Khymyn2016,Lebrun2018,Jani2021}. Among the various AFMs studied in recent decades, NiO and CoO thin films stand out due to their well-established high quality epitaxial growth, extensively explored spin structures \cite{Altieri2003,Arenholz2007, Li2011,Li2015,Li2016a,Yang2018,Xu2019,Xu2020a}, as well as their unique magnetic properties that are promising for spintronic applications \cite{Khymyn2017,Yang2022,Lee2021,Qiu2020,Dabrowski2020,Schmitt2024}.

Here, we utilize the magneto-optical birefringence (MOB) effect \cite{Xu2019,Xu2020a}, also known as the Sch{\"a}fer-Hubert effect \cite{Schaefer1990,Zhou2023}, to directly image the AFM domains in thin CoO and NiO films and to demonstrate how these domains can be manipulated by ultrafast laser pulses. In the case of a static laser beam, AFM domains undergo thermal demagnetization, as indicated by the formation of smaller, randomly distributed domains. Furthermore, we demonstrate that partial domain switching can be achieved by sweeping the laser beam across the sample parallel to one of the two possible N\'eel vector orientations, whereby the resulting temperature gradient generates a spatial variation in magnetic energy that drives domain wall motion.

\section*{Results}

\subsection{Imaging antiferromagnetic domains via magneto-optical birefringence \newline}
AFM domain images were acquired in reflection geometry using wide-field Kerr microscopy (WFKM) with a white-light LED source. The LED light, incident normally on the sample, was polarized at 45$^{\circ}$ relative to the in-plane component of the N\'eel vector \textbf{n} of the AFM films. To obtain MOB contrast, two images $I(+\theta)$ and $I(-\theta)$ were recorded with the analyzer (or quarter-wave plate) set symmetrically about extinction, yielding opposite contrast. The MOB signal was extracted from the asymmetry $I_{as}=\frac{I(+\theta)-I(-\theta)}{I(+\theta)+I(-\theta)}$ which enhances sensitivity to the in-plane projection of the N\'eel vector while suppressing artifacts from surface morphology \cite{Xu2019,Zhou2023}. An optical pump beam at 1035\,nm with tunable pulse duration, variable polarization and repetition rate was incident at 45$^{\circ}$ to the sample plane and focused to a 90\,$\mu$m diameter spot (intensity at $1/e^2$).

The single-crystalline NiO and CoO films were grown on MgO(001) substrates in an ultrahigh-vacuum system by molecular beam epitaxy (MBE) \cite{Zhu2014a,Li2015a,Xu2019,Xu2020a}. In NiO/MgO(001), tensile strain favors an out-of-plane orientation of the AFM spins; that is, they rotate away from the $\langle 11\bar{2} \rangle$ directions characteristic of bulk NiO toward the [001] axis \cite{Alders1998,Kim2010}. The critical thickness for lattice relaxation of NiO on MgO(001) is expected to be around 20\,nm \cite{Altieri2003}. However, a substantial canting of the N\'eel vector \textbf{n} away from the sample normal is already observed at smaller thicknesses; for a 10\,nm thick NiO film, the canting angle has been estimated to be approximately 10$^\circ$ away from the [001] direction \cite{Xu2019}. In NiO/MgO(001), four different twin domains (T domains) are present, while only one spin domain (S domain) with the largest out-of-plane component is energetically favored \cite{Altieri2003,Xu2019,Schmitt2023}. In contrast, for CoO/MgO(001), compressive strain induces a fourfold in-plane magnetic anisotropy, with an out-of-plane hard axis along [001] and two easy axes within the (001) plane along the $[110]$ and $[1\bar{1}0]$ directions \cite{Zhu2014a,Zhu2014b,Baldrati2020,Xu2020a}. Similarly to previous studies \cite{Xu2019,Xu2020a,Meer2023,Schmitt2024} both NiO and CoO films reveal excellent epitaxial growth with the lattice relation NiO[100](001)$||$MgO[100](001) and CoO[100](001)$||$MgO[100](001). The magnetic order is expected to be fully compensated in both AFMs. When NiO and CoO layers are grown together on MgO(001), the AFM spin structures of the individual layers can be further tuned via interfacial exchange coupling \cite{Zhu2014a}. In particular, a spin reorientation transition (SRT) of the NiO spins from an out-of-plane to an in-plane orientation can be induced by increasing the thickness, while the N\'eel temperature ($T_{\mathrm{N}}$) of the CoO layer can be significantly enhanced \cite{Zhu2014a}. Taking these considerations into account, AFM domains with different orientations and $T_{\mathrm{N}}$ values can be engineered through an appropriate combination of NiO and CoO thin films \cite{Zhu2014a}.

\subsection{Thermal demagnetization with a static laser beam \newline} 

Optical manipulation of magnetic domains usually involves an ultrafast demagnetization \cite{Beaurepaire1996}, in which pumping with laser pulses results in a strong non-equilibrium excitation of the material’s electron system, which subsequently transfers its energy to the spin system. However, given the photon energies available in the visible and infrared spectrum typically obtained from ultrafast lasers, this excitation scheme is not so easily realized for antiferromagnetic NiO and CoO, which are insulators with bandgaps of at least a few eV. To overcome this limitation, an adjacent metallic layer is used, so that an ultrashort laser pulse with photon energy below the bandgap of the insulating AFM can excite the hot electron system of the metallic layer, which then transfers energy to the spin system of the AFM \cite{Wust2022,Meer2023,Zhao2025}. Our results demonstrate that the domain structures of both CoO and NiO films can be modified by exposure to optical pulses, as long as they are capped with metallic layers. For NiO and CoO films capped with non-metallic layers, such as Al$_2$O$_3$, we have not observed any effects due to optical pumping (see Supplementary Fig. S1). Here we focus on AFM films with a nonmagnetic, metallic Pt (2\,nm) capping layer, which effectively protects against oxidation without significantly diminishing the intensity of light reaching the AFM layer, or the MOB effect \cite{Meer2023,Schmitt2024}.

\begin{figure} 
\centering
\includegraphics[width=1.0\linewidth]{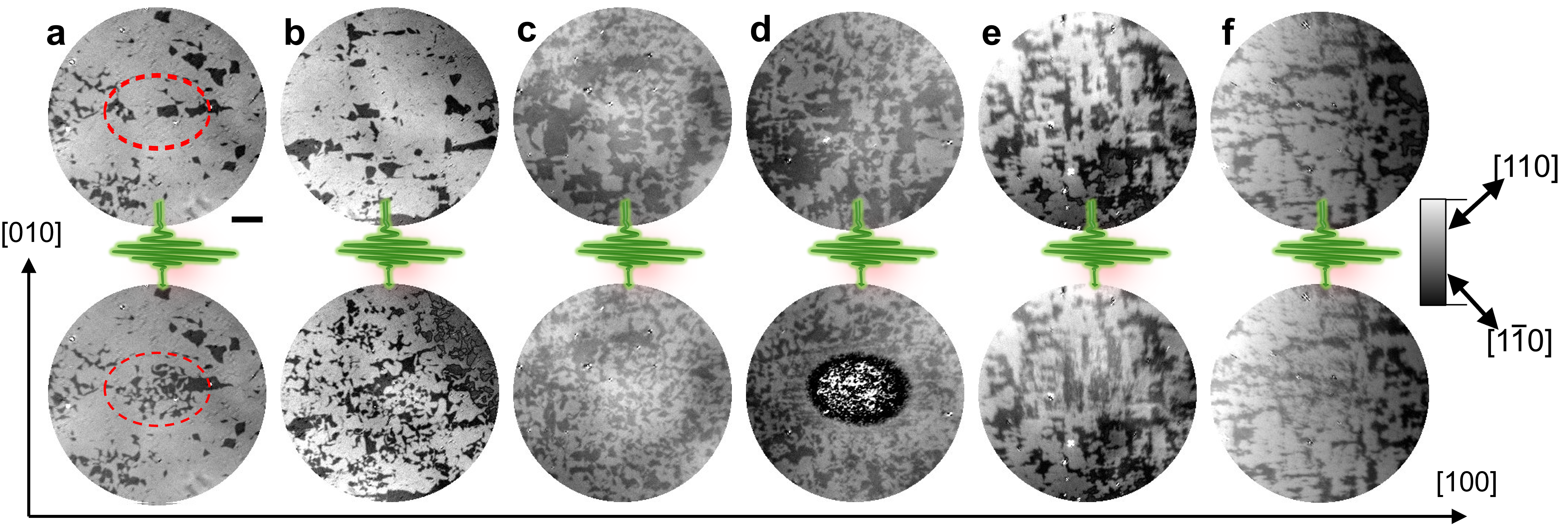}
  \caption{Domain structure images of thin CoO and NiO films before (top row) and after (bottom row) optical pumping with p-pol, 1\,ps laser pulses with photon energy E = 1.2\,eV. \textbf{CoO(8\,nm)/Pt(2\,nm)/Al$_2$O$_3$(3\,nm)} with single pulse excitaiton at fluence (\textbf{a}) $F$\,=\,0.6\,mJ/$\mathrm{cm^2}$ and (\textbf{b}) $F$\,=\,11\,mJ/$\mathrm{cm^2}$. \textbf{CoO(8\,nm)/Pt(2\,nm)} with excitation by (\textbf{c}) a single pulse at $F$\,=\,11\,mJ/$\mathrm{cm^2}$ and (\textbf{d}) 10$^6$ pulses at $F$\,=\,11\,mJ/$\mathrm{cm^2}$. \textbf{NiO(16\,nm)/Pt(2\,nm)} for single pulse excitation at (\textbf{e}) $F$\,=\,11\,mJ/$\mathrm{cm^2}$. \textbf{CoO(8\,nm)/NiO(5\,nm)/Pt(2\,nm)} for single pulse excitaiton at (\textbf{f}) $F$\,=\,11\,mJ/$\mathrm{cm^2}$. The images were acquired by WFKM at RT with the laser light polarized along the [100] axis. Bright and dark areas correspond to the in-plane projections of the N\'eel vector orientations \textbf{n} along the $[110]$ and $[1\bar{1}0]$ axes, respectively. The scale bar is 10 $\mu$m in length.} 
  \label{fig1}
\end{figure}

The top row of Fig.\ref{fig1} shows images of the AFM domain structures in the ground (virgin) state acquired by WFKM for four different samples: CoO(8\,nm)/Pt(2\,nm)/Al$_2$O$_3$(3\,nm) [Fig.\ref{fig1}(a,b)], CoO(8\,nm)/Pt(2\,nm) [Fig.\ref{fig1}(c,d)], NiO(16\,nm)/Pt(2\,nm) [Fig.\ref{fig1}(e)], and \break CoO(8\,nm)/NiO(5\,nm)/Pt(2\,nm) [Fig.\ref{fig1}(f)]. In all cases, we probe the in-plane projections of AFM spins along the $[110]$ and $[1\bar{1}0]$ axes, corresponding to bright and dark areas, respectively. The relationship between the direction of \textbf{n} and the optical contrast was characterized in the CoO/Fe film, where the orthogonal interfacial coupling between the Fe spins and the CoO AFM spins allows the orientation of \textbf{n} to be controlled with the in-plane magnetic field \cite{Xu2020a}. The domain walls appear to be preferentially oriented along the $[010]$ and $[100]$ axes. 

The bottom row of Fig.~\ref{fig1} illustrates the effect of optical pumping for different fluences $F$ and pulse numbers. Laser exposure results in smaller, randomly shaped domains without preferential elongation, while spins remain aligned along the $[110]$ and $[1\bar{1}0]$ axes. This behavior arises from magnetoelastic coupling to the substrate \cite{Wittmann2022,Meer2022}, which favors a multidomain state with a nearly homogeneous domain distribution. The as-grown domain structure is nonequilibrium in terms of domain size and distribution; laser heating promotes domain wall motion and N\'eel vector reorientation, enabling relaxation toward an equilibrium state characterized by more uniformly distributed, submicron domains. Starting with the CoO(8\,nm)/Pt(2\,nm)/Al$_2$O$_3$(3\,nm) sample exposed to a single pulse, one can observe that the AFM domains can be modified locally at the center of the field of view (FOV) at low fluence $F$\,=\,0.6\,mJ/$\mathrm{cm^2}$ [as indicated by a dashed red ellipse in Fig. \ref{fig1}(a)]. The size of the modified region scales with fluence, and for $F$\,=\,11\,mJ/$\mathrm{cm^2}$ [Fig. \ref{fig1}(b)], nearly all domains within the FOV are affected. The effect of the optical pumping is very similar for the sample CoO(8\,nm)/Pt(2\,nm) [Fig. \ref{fig1}(c)]. Figure \ref{fig1}(d) demonstrates that exposure to 10$^6$ pulses at $F$\,=\,11\,mJ/$\mathrm{cm^2}$ results in permanent damage to the sample. Note however that, in general, for lower fluence, both single pulses and multipulses result in essentially similar modification of the AFM domains (Supplementary Fig. S2). For samples containing NiO films: NiO(16\,nm)/Pt(2\,nm) [Fig. \ref{fig1}(e)] and CoO(8\,nm)/NiO(5\,nm)/Pt(2\,nm) [Fig. \ref{fig1}(f)], the optical pumping results in creation of even smaller domains, with some features that can hardly be resolved with the current optical imaging method. Furthermore, the modified area of the samples with NiO is reduced as compared to those with only CoO for the same fluence. This can be attributed to the different N\'eel temperatures, which are expected to be around 330\,K for CoO(8\,nm) \cite{Xu2020a}, and around 500\,K and 515\,K for CoO(8\,nm)/NiO(5\,nm) and NiO(16\,nm), respectively \cite{Xu2019,Zhu2014a}.

\subsection{Domain switching induced by a swept laser beam \newline}

\begin{figure} 
\centering
\includegraphics[width=0.45\linewidth]{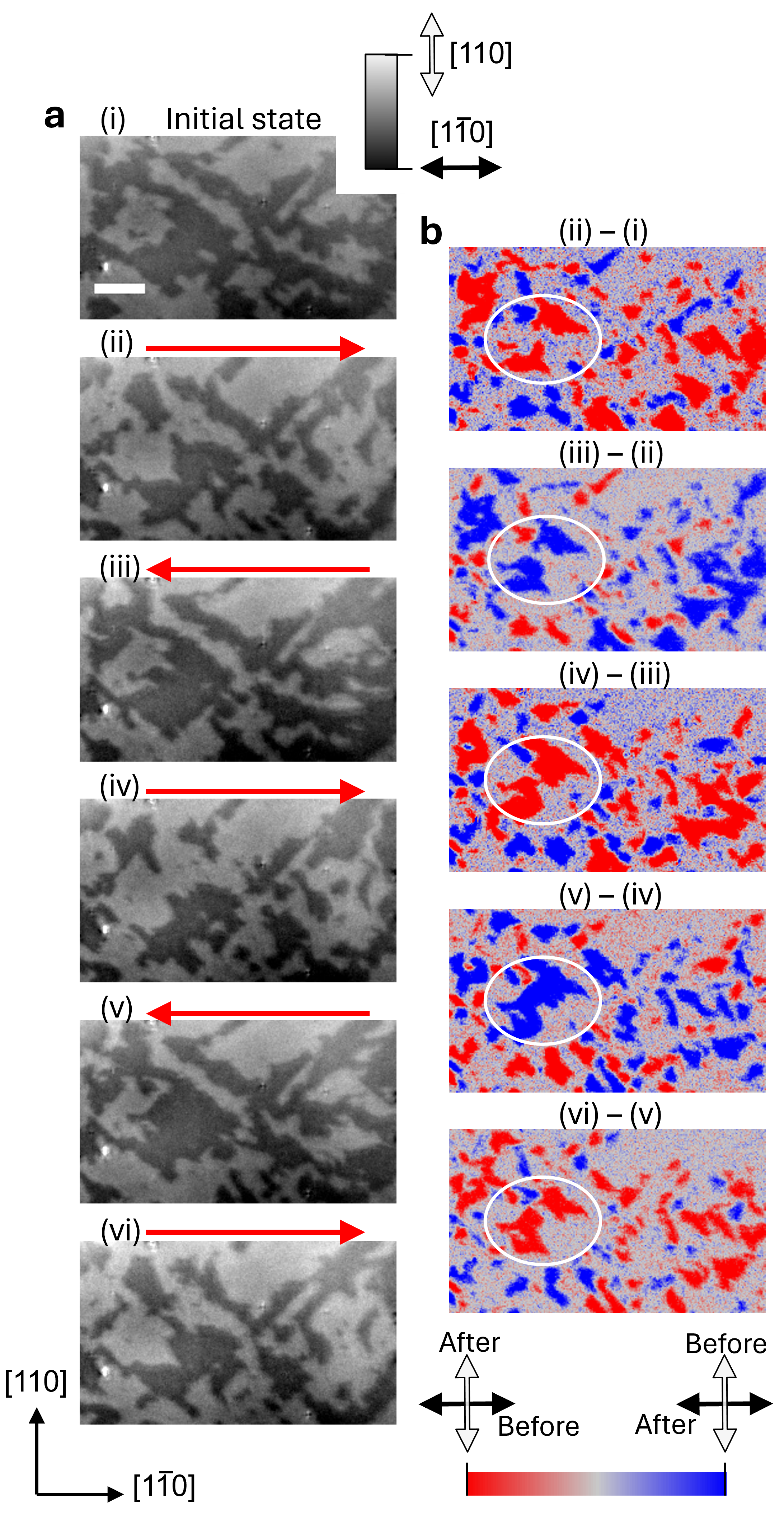}
   \caption{(\textbf{a}) Domain structure images  acquired after sweeping the laser beam across the surface at a velocity of $v$\,=\,100\,$\mu$m/s along the $[1\bar{1}0]$ axis, as indicated by the red arrows above the images. (\textbf{b}) Differential images [\textbf{(ii)$-$(i)} etc.] show that some of the domains can be switched back and forth (see e.g. the domains marked by the white ellipse). Red and blue contrast in the differential images correspond to switching from $[1\bar{1}0]$ to $[110]$, and from $[110]$ and $[1\bar{1}0]$, respectively, while intermediate gray contrast indicates areas where no switching occurs. The measurements were performed at RT for the CoO(8\,nm)/Pt(2\,nm) sample, with p-pol light, $F$\,=\,2.3\,mJ/$\mathrm{cm^2}$, 1\,MHz repetition rate and 8\,ps pulse duration. The scale bar is 10 $\mu$m in length.}
  \label{fig2}
\end{figure}

Although a static beam can lead to a local rearrangement of AFM domains, no controlled or reversible manipulation of the N\'eel vector orientation was achieved for any of the samples studied within the available parameter space. However, previous studies on all-optical switching (AOS) in ferromagnetic and ferrimagnetic materials have shown that sweeping the laser beam across the sample surface can facilitate AOS. This is attributed to the additional thermal gradient, which helps activate domain wall depinning and promotes domain wall motion \cite{Mangin2014,Lambert2014,Hadri2016a,John2017,Hadri2017,Hees2020,Dabrowski2021}. To this end, we performed experiments in which the laser beam was swept across the sample surface. Various samples from Fig. \ref{fig1} were tested for different optical pumping parameters (see Supplementary Figs. S3 and S4 for exemplary images). The most substantial changes in domain structure were observed for CoO films when the beam was swept parallel to one the easy axes of the N\'eel vector \textbf{n}. Figure \ref{fig2} shows the resulting changes in the CoO domain structure after sweeping the beam back and forth along the $[1\bar{1}0]$ axis, as indicated by red arrows. Histograms of the intensity distribution after consecutive sweeps indicate that the switching is also manifested in the overall preferential orientation of the domains (see Supplementary Fig. S5). The changes to the domain structure are clearly visible in the differential images between consecutive frames [(ii)$-$(i), (iii)$-$(ii) etc.] shown in Fig.\ref{fig2}(b). Analysis of the consecutive differential images reveals that many AFM domains undergo 90$^{\circ}$ switching in a reproducible manner (see, for example, the area marked by the white ellipse).  Interestingly, both types of domains, with $\mathbf{n}\parallel [110]$ and $\mathbf{n}\parallel [1\bar{1}0]$ can be switched simultaneously during a single sweeping event. See Supplementary Fig. S6 for enlarged (zoomed-in) views of the areas marked by the white ellipse.

\subsection{Thermal gradient and the ponderomotive force  \newline}

We examine the physical origin of the laser-induced partial switching of AFM domains in CoO/Pt. The effect is found to be independent of the pump beam polarization and is absent under stationary beam conditions or single-pulse excitation. Additionally, the observed optically induced changes, including those observed with a stationary beam, are cumulative and depend on both the number of pulses and the repetition rate. These observations suggest that the effect is thermal in nature, allowing us to exclude non-thermal mechanisms such as the inverse Faraday effect \cite{Kimel2004,Tzschaschel2017} and the inverse Cotton-Mouton effect \cite{Kalashnikova2007,Tzschaschel2017}. Laser-induced heating leads to partial demagnetization, which can be described as a reduction in saturation magnetization ($M_s$) and the corresponding magnetic energy density ($w_\mathrm{mag}\sim M_s^2$). The laser beam creates an inhomogeneous temperature distribution $T(\mathbf{r})$, resulting in a spatially varying magnetic energy $w_\mathrm{mag}[T(\mathbf{r})]$ that follows the beam's intensity profile. As a result, regions at higher temperature possess lower magnetic energy and are energetically favored, irrespective of the local orientation of the N\'eel vector. The spatial gradient of magnetic energy across a domain wall gives rise to a driving pressure on the wall that originates from the difference in magnetic energy density on its two sides. This pressure can be understood within the general concept of a ponderomotive force $\mathbf{F}_\mathrm{pond}$, which describes motion induced by gradients in energy density.  While ponderomotive forces are often discussed in the context of charged particles in inhomogeneous electromagnetic fields, the concept also applies to magnetic domain walls subject to spatial variations of magnetic energy density \cite{BRECHET2013,Gomonay2024}. 

\begin{figure} 
\centering
\includegraphics[width=0.6\linewidth]{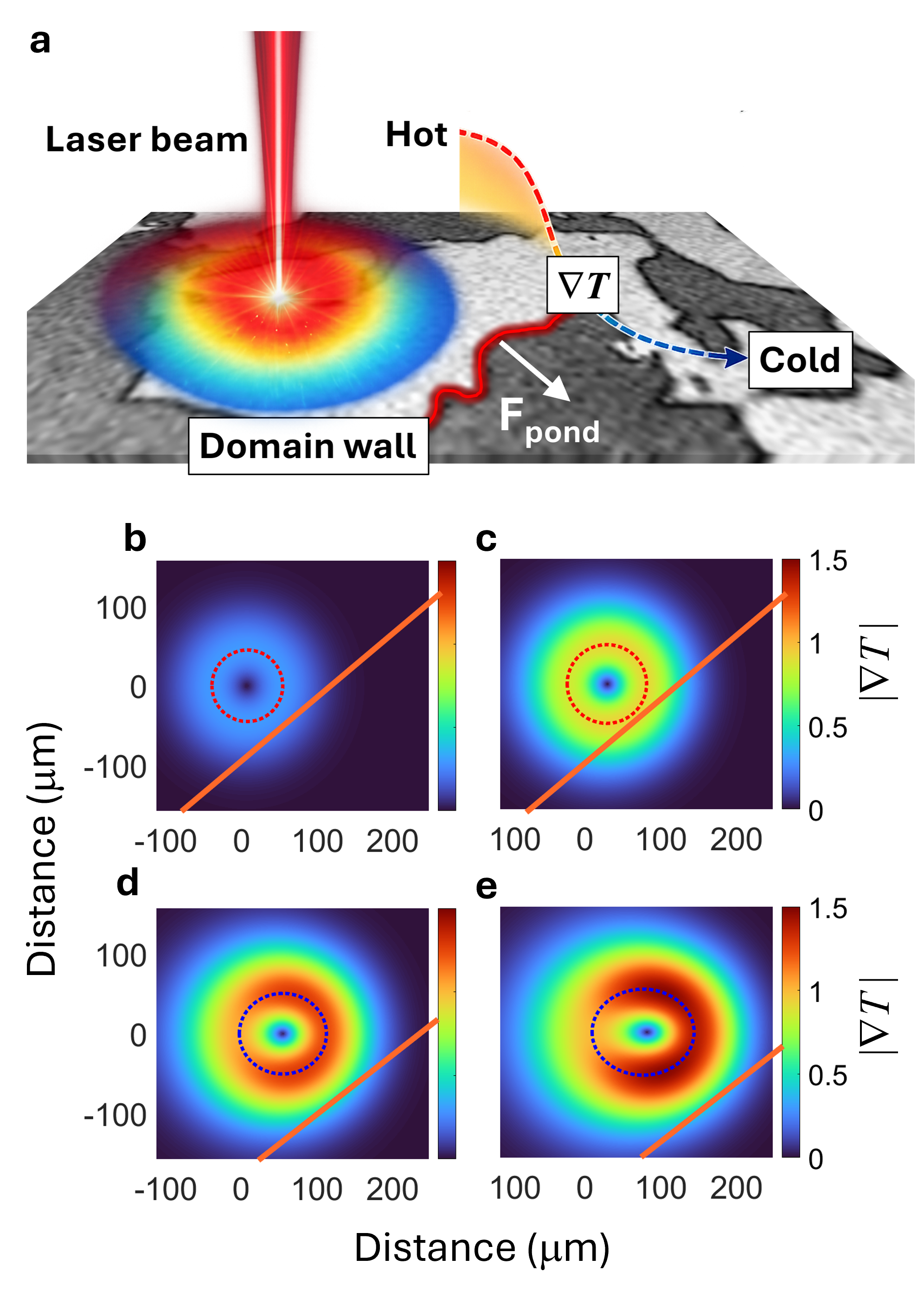}
   \caption{(\textbf{a}) Schematic illustration showing the thermal gradient induced by the laser beam, and associated with it the ponderomotive force $\mathbf{F}_\mathrm{pond}$, exerting the pressure on the domain wall wall (marked in red). The domain structure image has been obtained from the CoO(8\,nm)/Pt(2\,nm) sample.(\textbf{b-e}) Snapshots of the calculated temperature gradient $\nabla T$ induced by the laser beam as it moves parallel to the N\'eel vector along the [110] axis. The orange line indicates the position of the domain wall, which is parallel to the [100] axis (shown for illustration purposes only). At (\textbf{b})  $t=0.12\tau^\prime$, $\mathbf{F}_\mathrm{pond}$ at the location of the domain wall is below threshold and the domain wall does not move; (\textbf{c}) $t=0.48\tau^\prime$, (\textbf{d}) $t=0.83\tau^\prime$, and (\textbf{e}) $t=1.2\tau^\prime$ the $\mathbf{F}_\mathrm{pond}$ at the location of the domain wall is above the critical value due to the motion of the beam and the domain wall moves with steady average velocity. 
   } 
  \label{fig3}
\end{figure}

The width of the domain wall ($x_\mathrm{DW}$) is of order 100 nm and is small compared to the characteristic scale of the temperature gradient (of order tens of $\mu$m). Under this assumption, the value of the ponderomotive force acting at the domain wall is given by $\mathbf{F}_\mathrm{pond}=-\nabla w_\mathrm{mag}[T(\mathbf{r})]$, and the force is directed toward the colder domain. Although the ponderomotive force can push the domain wall, several factors prevent observable modification of the domain structure by a static beam. In particular, the domain walls are pinned due to an inhomogeneous distribution of spontaneous strains and crystal defects. Therefore, to set the domain wall in motion, the ponderomotive force must overcome the pinning barrier. This occurs at the point of maximum temperature gradient. The idea is schematically illustrated in Fig. \ref{fig3}(a). Once depinned, the domain wall moves outside the laser spot to a colder region where the value of the ponderomotive force is lower. A further reduction in the ponderomotive force is related to heat diffusion and temperature relaxation following the laser pulse. Combined with damping, this results in finite displacement of the domain wall. In conclusion, a static beam cannot support the steady motion of the domain wall, in agreement with our experimental observations.

In contrast, the moving beam can drag the domain wall through noticeable distances, thereby inducing switching between the domains.  When the velocities of the beam and the domain wall are comparable, the moving domain wall experiences a ponderomotive force that is sufficiently large to compensate for damping and allow the domain wall to travel a greater distance before being stopped by other pinning defects (e.g., the junction of domain walls). Figures \ref{fig3}(b)-(e) present calculated temperature gradient colormaps for the laser beam moving parallel to the N\'eel vector along the [110] axis at different times $\tau^\prime\equiv \sigma^2C_V/(2\kappa)$, where $C_V$ is a heat capacity, $\kappa$ is a heat conductivity, $\sigma$ is the radius of the beam. The orange line (shown for illustrative purposes only) schematically marks the position of the domain wall, oriented along the [100] axis.

\begin{figure} 
\centering
\includegraphics[width=0.6\linewidth]{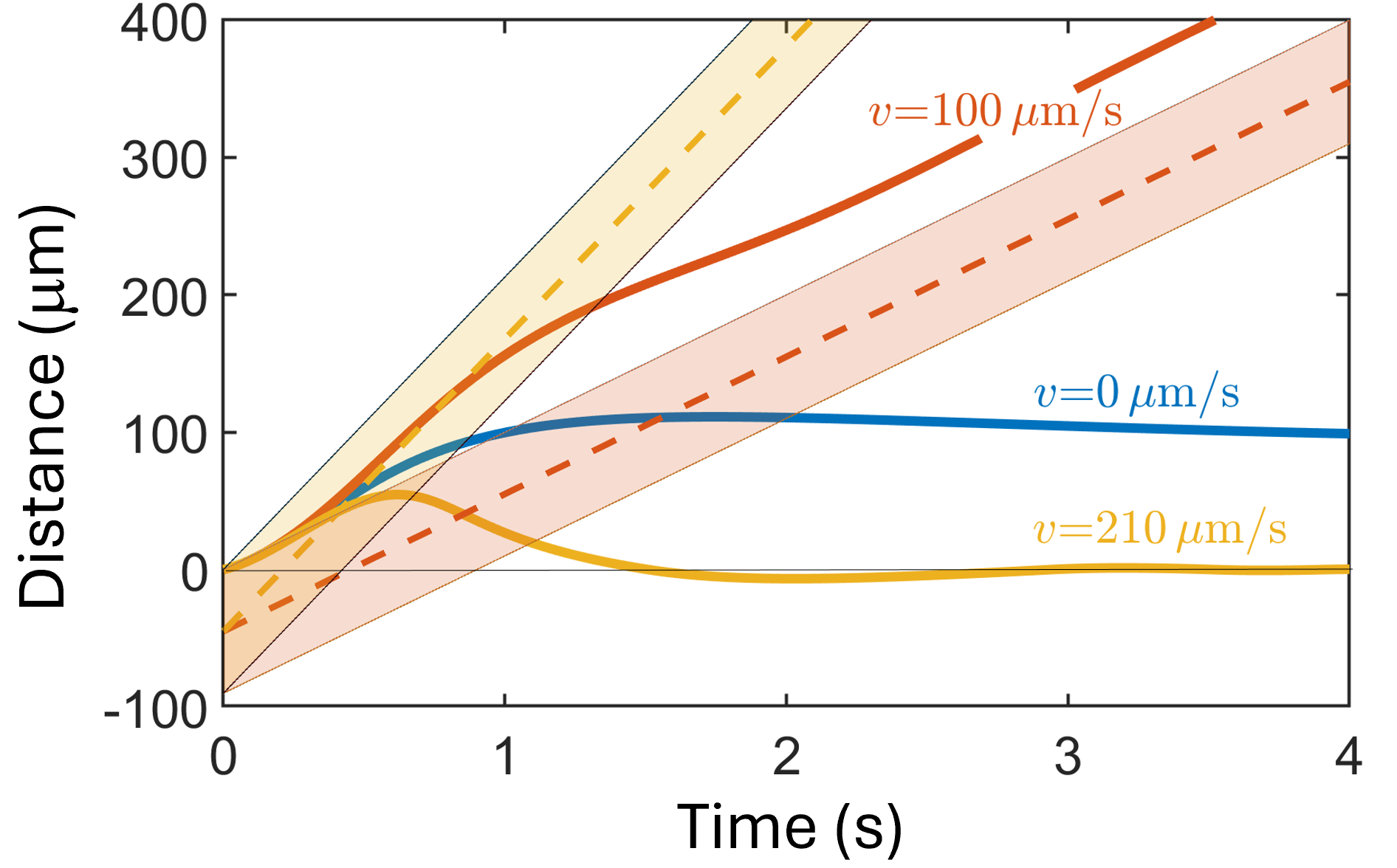}
   \caption{(\textbf{a}) Displacement of the domain wall (along the coordinate ${X}$) as a function of time for a static ($v=0$) and moving beams ($v$\,=\,100\,$\mu$m/s and $v$\,=\,210\,$\mu$m/s). Dashed lines show the displacement of the beam center (also along ${X}$, parallel to the domain wall motion), and the shaded regions represent the width of the laser spot.
   } 
  \label{fig4}
\end{figure}

To gain a better understanding of the switching mechanism, we modeled the light-induced motion of a single domain wall. We used the Thiele-like equation for the domain wall center, \textit{X} (see Supplementary Sections S2, S3 and S5 for the details): 
\begin{equation}\label{eq_force_v2}
    \ddot{X}+\frac{c^2}{\tau x_\mathrm{DW}^2}\dot{X}=F_\mathrm{pin}(X)+F_\mathrm{pond}(t,X),
\end{equation}
where $c$ is the limiting magnon velocity, $x_\mathrm{DW}$ is the domain wall width, $\tau$ is the relaxation time (caused by the Gilbert damping). The pinning force $F_\mathrm{pin}(X)$ is calculated by assuming that the pinning has magnetoelastic origin: 
\begin{equation}\label{eq_pining_force}
    F_\mathrm{pin}(X)\equiv-\frac{c^2\mathrm{sign}(X)}{2\sqrt{2}x_\mathrm{DW}}\left[\frac{1}{2}-\frac{1}{\cosh^2[(|X|+\xi_0)/x_\mathrm{DW}]}\right],
\end{equation}
where $\xi_0=x_\mathrm{DW}\sinh^{-1}1$. The ponderomotive force 
\begin{equation}\label{eq_ponderomotive}
    F_\mathrm{pond}(t,X)=-\sqrt{2}\frac{c^2}{M_s}\frac{\partial M_s}{\partial T}\partial_\xi T(t,\mathbf{r})
\end{equation}
is calculated by assuming a temperature distribution induced by a laser beam with Gaussian intensity distribution.  An example of $F_\mathrm{pond}(t,\mathbf{r})$ caused by the moving beam is shown in Fig. \ref{fig4}. For simplicity we assumed that $\partial M_s/\partial T=const$. We find that steady domain wall motion occurs when the velocity of the laser beam is comparable to that of the domain wall (red curve, $v$\,=\,100\,$\mu$m/s in Fig. \ref{fig4}). If the beam moves too quickly, the wall enters a region where the ponderomotive force reverses direction and the wall returns to its initial position (yellow curve, $v$\,=\,210\,$\mu$m/s in Fig. \ref{fig4}). When the beam moves too slowly or remains stationary, the wall reaches a region of negligible ponderomotive force and stops (blue curve, $v$\,=\,0\,$\mu$m/s in Fig. \ref{fig4}). This model can thus guide optimization of switching performance by adjusting the beam sweep velocity.

We also note that the CoO domain structure is complex, consisting of a network of intersecting domain walls oriented along the hard magnetic axes $[100]$ and $[010]$ due to magnetoelastic effects. The most efficient way to switch the domain structure and overcome pinning at the junctions is to apply a force that sets both types of intersecting wall into motion. This can be achieved by applying a ponderomotive force along $[110]$ or $[1\bar{1}0]$ i.e., along the easy magnetic axes at 45$^\circ$ with respect to the domain walls, as observed experimentally (Fig. \ref{fig2}) (see Supplementary Section S4 for more details).

\section*{Discussion}

In summary, we identify a new mechanism for optical manipulation of fully compensated insulating AFM domains. This approach requires neither magnetic nor electric fields, operates in continuous thin films, and paves the way toward all-optical antiferromagnetic recording using laser pulses. We show that a metallic Pt capping layer enables efficient absorption of photon energies well below the bandgap of NiO and CoO, leading to thermal demagnetization down to a single laser pulse. Sweeping the beam across the sample surface along the N\'eel vector generates a temperature gradient that drives domain-wall motion toward colder regions and enables partial 90$^{\circ}$ switching of AFM domains. Previous studies on current-induced switching in CoO/Pt have shown that the primary driving mechanism is the thermomagnetoelastic effect \cite{Baldrati2020,Schmitt2024,Wu2024a}, where the final orientation of the N\'eel vector \textbf{n} is either perpendicular ($\textbf{n}\perp\textbf{j}$) or parallel ($\textbf{n}\parallel\textbf{j}$) to the current density \textbf{j} \cite{Wu2024a}. In contrast, the optically induced switching observed here is fundamentally different: the same AFM domains can be switched back and forth by 90$^{\circ}$ simply by reversing the direction of the moving laser beam, without changing beam polarization or fluence. The observed switching occurs only in localized regions, indicating that domain wall pinning plays a key role. Improved understanding and control of these pinning sites should enhance the reproducibility and precision of switching driven by the ponderomotive force proposed here.

\clearpage
\noindent
\large{\textbf{Methods}}

\normalsize

\medskip
\noindent
\textbf{Domain structure imaging\newline} Antiferromagnetic domains were imaged via magneto-optical birefringence (MOB) using wide-field Kerr microscopy (WFKM). The LED light, incident normally on the sample, was linearly polarized at 45$^{\circ}$ relative to the in-plane component of the N\'eel vector \textbf{n} of the AFM films. To obtain MOB contrast, two images $I(+\theta)$ and $I(-\theta)$ were recorded with the analyzer (or quarter-wave plate) set symmetrically about extinction, yielding opposite contrast. The MOB signal was extracted from the asymmetry $I_{as}=\frac{I(+\theta)-I(-\theta)}{I(+\theta)+I(-\theta)}$ which enhances sensitivity to the in-plane projection of the N\'eel vector while suppressing artifacts from surface morphology. For optically induced magnetization changes, the fundamental laser output at 1035 nm with variable pulse duration was used as an optical pump, with p-polarized light incident at 45$^{\circ}$ to the sample plane and focused to a 90 $\mu$m diameter spot (intensity at $1/e^2$) \cite{Dabrowski2021}. The number of pulses was controlled via an external trigger signal (a TTL pulse of specified width) generated by a function generator. All measurements were conducted at room temperature. 


\subsection{Data Availability\newline} 
All the data that support the findings of this study have been included in the main text and Supplementary Information. 



\medskip
\noindent
\textbf{Acknowledgements\newline}
 M.D., C.S., P.S.K. and R.J.H. acknowledge the support of the Engineering and Physical Sciences Research Council (EPSRC) through grants EP/W006006/1, EP/V048538/1, EP/R008809/1 and EP/V054112/1. Y.W. acknowledges the support of the National Key Research and Development Program of China (Grant No. 2024FYA1408500) and the National Natural Science Foundation of China (Grant No. 12434003, and No. 12221004). Measurements were performed in the Exeter Time-Resolved Magnetism Facility (EXTREMAG) facility funded by grants EP/R008809/1 and EP/V054112/1. O. G. acknowledges funding by the DFG Grant No. TRR 288-422213477 (project A12) and TRR 173-268565370 (projects A11 and B15).

\medskip
\noindent
\textbf{Author contributions\newline} 
M.D. conceived the idea, performed measurements and analyzed the data. T.W., J.X. and Y.W. fabricated and characterized the samples. O.G. developed the theoretical model and contributed to the interpretation. C.R.J.S., P.S.K., Y.W. and R.J.H. gave suggestions on the experiments. M.D. prepared the original manuscript with help from R.J.H. and O.G. All authors discussed the results and contributed to the manuscript.

\medskip
\noindent
\textbf{Competing interests\newline}
The Authors declare no Competing Financial or Non-Financial Interests. 
\clearpage

\end{document}